\documentclass[aps,prb,showpacs,twocolumn,reprint,superscriptaddress]{revtex4-2}
\usepackage{amsmath}
\usepackage{amssymb}
\usepackage{graphicx}
\usepackage{color}
\usepackage{multirow}
\usepackage{dcolumn}
\usepackage{textcomp}
\usepackage{longtable}
\usepackage{makecell}
\usepackage{soul, color, xcolor}

\begin{document}

\title{The ground state of CuInP$_2$S$_6$ thin films: A study of the deep potential method}

\author {Shengxian Li}
\affiliation{Key Laboratory for Matter Microstructure and Function of Hunan Province, Key Laboratory of Low-Dimensional Quantum Structures and Quantum Control of Ministry of Education, Changsha 410081, China}
\affiliation{Hunan Research Center of the Basic Discipline for Quantum Effects and Quantum Technologies, Changsha 410081, China}
\affiliation{School of Physics and Electronics, Hunan Normal University, Changsha 410081, China}

\author {Jiaren Yuan}
\affiliation{School of Physics and Materials Science, Nanchang University, Nanchang, Jiangxi, 330031, People’s Republic of China }

\author{Tao Ouyang}
\affiliation{Hunan Key Laboratory for Micro-Nano Energy Materials and Device and School of Physics and Optoelectronics, Xiangtan University, Xiangtan 411105, Hunan, China}

\author{Anlian Pan}
\affiliation{Key Laboratory for Matter Microstructure and Function of Hunan Province, Key Laboratory of Low-Dimensional Quantum Structures and Quantum Control of Ministry of Education, Changsha 410081, China}
\affiliation{Hunan Research Center of the Basic Discipline for Quantum Effects and Quantum Technologies, Changsha 410081, China}
\affiliation{School of Physics and Electronics, Hunan Normal University, Changsha 410081, China}
\affiliation{Key Laboratory for Micro-Nano Physics and Technology of Hunan Province, 
College of Materials Science and Engineering, Hunan University, Changsha 410082, China}

\author {Mingxing Chen}
\email{mxchen@hunnu.edu.cn}
\affiliation{Key Laboratory for Matter Microstructure and Function of Hunan Province, Key Laboratory of Low-Dimensional Quantum Structures and Quantum Control of Ministry of Education, Changsha 410081, China}
\affiliation{Hunan Research Center of the Basic Discipline for Quantum Effects and Quantum Technologies, Changsha 410081, China}
\affiliation{School of Physics and Electronics, Hunan Normal University, Changsha 410081, China}

\date{\today}

\begin{abstract}
The two-dimensional ferroelectric (FE) material CuInP$_2$S$_6$ (CIPS) has garnered considerable interest due to its out-of-plane ferroelectricity at room temperature. However, a notable discrepancy exists between experiments and density functional theory (DFT) calculations regarding the ground state of CIPS thin films: experiments suggest a state with net polarization, while DFT predicts an antiferroelectric (AFE) state as the lowest-energy state. Here, we investigate the stability of polarization states in CIPS thin films by combining first-principles calculations with the deep potential (DP) method. Our results reveal that for films thicker than the bilayer, an AFE state that has intralayer AFE ordering in the inner layers and intralayer FE ordering in the two surface layers has the lowest electronic energy. This state is significantly lower than the uniform FE state. In addition, we find that a ferrielectric (FiE) state with pure intralayer FE ordering is very close to the AFE state in energy. By using the DP model, we calculated the phonon free energy of CIPS thin films. For the monolayer, the intralayer FE ordering possesses a lower phonon free energy than the intralayer AFE ordering. This energetic preference for the intralayer FE ordering maintains as the thickness grows. Consequently, when the phonon-free energy is incorporated, the FiE state becomes energetically favorable over the AFE state for multilayers CIPS. Our findings demonstrate that the inclusion of vibrational entropy stabilizes the FiE state as the ground state in multilayers CIPS at finite temperatures, reconciling the previous discrepancy between experimental observations and DFT predictions. This insight is vital for understanding the FE properties of CIPS and its potential applications in devices.
\end{abstract}

\maketitle

\section{Introduction}
Ferroelectric (FE) materials have promising applications in non-volatile memory devices with ultra-high storage density and low power consumption~\cite{muralt2000ferroelectric,setter2006ferroelectric,catalan2009physics,martin2016thin,vaz2021epitaxial,scott1989ferroelectric,lanza2022memristive,wang2021two,si2019ferroelectric,chanthbouala2012solid,tian2016tunnel,wen2013ferroelectric,yang2019ferroelectric}. The discovery of two-dimensional FE materials such as CuInP$_2$S$_6$ (CIPS)~\cite{maisonneuve1997ferrielectric,belianinov2015cuinp2s6,liu2016room}, In$_2$Se$_3$~\cite{ding2017prediction,zhou2017out,cui2018intercorrelated}, and SnTe~\cite{fei2016ferroelectricity,chang2016discovery} favors miniaturization of the devices. Among the 2D ferroelectrics, CIPS has attracted much attention due to its stable out-of-plane ferroelectricity at room temperature~\cite{maisonneuve1997ferrielectric,belianinov2015cuinp2s6,liu2016room,sivadas2023scale,io2022strong,deng2020thickness,liu2024reversible}. In addition, it exhibits various exotic properties, including tunable quadruple-well polarization states~\cite{brehm2020tunable,neumayer2020alignment}, negative longitudinal piezoelectricity~\cite{you2019origin,qi2021widespread}, and the bulk photovoltaic effect~\cite{li2021enhanced,yu2024flexoelectric}. In particular, a recent study found that CIPS thin films can be controlled in double, quadruple, and sextuple polarization states~\cite{li2024realization}. 

In the CIPS crystal, each monolayer consists of a sulfur framework with octahedral voids occupied by metal cations (Cu$^{\mathrm{I}}$, In$^{\mathrm{III}}$) and P-P pairs, which are arranged in triangular patterns. The emergence of out-of-plane ferroelectricity primarily stems from the antiparallel displacements of the two distinct cations relative to the layer's central plane~\cite{maisonneuve1997ferrielectric}. Experimental studies have confirmed the persistence of 2D ferroelectricity in CIPS films down to 4 nm thickness (corresponding to $\sim$ 6 CIPS layers)~\cite{liu2016room}. However, DFT calculations find that an intralayer AFE ordering has a much lower energy than the FE ordering for the monolayer~\cite{yu2021interface,reimers2018van,he2023unconventional,sun2019strain,tawfik2018van,yu2023polarization,li2024realization}. Whereas the bilayer tends to have intralayer FE orderings that are antiferroelectrically coupled between layers, i.e., interlayer AFE ordering. For the trilayer, a state involving the intralayer AFE ordering in the inner layer and intralayer FE ordering in the two surface layers is found to be the lowest energy state~\cite{yu2023polarization,li2024realization}. Recently, an interlayer AFE state similar to that for the bilayer was found to be more stable than the FE state for CIPS thin films~\cite{neumayer2022nanoscale}. This state displays the intralayer FE ordering across all layers, with the surface layer exhibiting antiparallel polarization relative to its adjacent sublayer, forming an interlayer AFE coupling near the surface. Moreover, DFT calculations also suggest that the interlayer AFE ordering was favored for CuInP$_2$Se$_6$ when its film thickness is reduced below six layers~\cite{song2017off}. Understanding the complexity of the polarization states in CIPS thin films and resolving the above discrepancy between experiments and DFT calculations are crucial for their further applications in devices. 

Given the hopping nature of the copper atoms in CIPS~\cite{maisonneuve1997ferrielectric}, it is necessary to evaluate the stability of the polarization states in this system by going beyond the use of standard DFT electronic energy, e.g., to include lattice dynamics for evaluation. However, because the CIPS films can contain hundreds of atoms, calculating the phonon dispersions of these systems using first-principles methods becomes infeasible. Recently, machine learning potential methods have demonstrated their ability to model large-scale atomic systems with high efficiency and precision~\cite{behler2007generalized,bartok2010gaussian,shapeev2016moment,zhang2018deep,fan2021neuroevolution}. For instance, the deep potential (DP) method was successfully used to evaluate the Curie temperatures of FE materials~\cite{he2023unconventional,he2022structural,wu2021deep,wu2021accurate}. Thus, applying this method to investigate the lattice dynamics of CIPS thin films may help resolve the debate regarding the ground state of the polarization in these systems. 

In this paper, we investigate the stability of polarization states in CIPS thin films by combining first-principles and the DP methods, allowing us to calculate the phonon free energy of CIPS more than 26 nm. We find that the phonon free energy of the intralayer FE state is lower than that of the intralayer AFE state. This relative stability is robust against the interlayer coupling and gets enhanced as temperature increases. Therefore, by including the contributions of phonons to the total energy, one can obtain a ferrielectric (FiE) ordering rather than the AFE or FE one as the ground state for CIPS multilayers. Furthermore, our molecular dynamics simulation also confirms this finding. This paper is organized as follows. We present the calculation methods and details in Sec.~\ref{method}.  Then we show the energy calculations of the polarization states for the monolayer and bilayer CIPS and the benchmarking of the DP model in Sec.~\ref{DP}. The results for multilayers CIPS are presented in Sec.~\ref{multilayers}. The contribution of phonon free energy on the stability of polarization states in CIPS thin films are discussed in Sec.~\ref{phonon}. Finally, the results are summarized in Sec.~\ref{conclusions}.

\section{Methods}
\label{method}
According to the DP method~\cite{zhang2018deep,zhang2018end}, the total energy \textit{E} of the system is considered to be the sum of the energy \textit{E$_i$} of each atom. Here, \textit{E$_i$} depends on its local environment \textit{R$_i$}, which contains the radial and angular information of other neighboring atoms within the cutoff radius \textit{R$_c$} of atom \textit{i}. To preserve the translational, rotational, and permutational symmetries of the system, the local environment \textit{R$_i$} is mapped to a descriptor \textit{D$_i$} by the embedding neural network. The parameters of the embedding neural network and fitting neural network are continuously optimized in the process of minimizing the loss function \textit{L$_p$}. The loss function \textit{L$_p$} is defined as \begin{equation}
L\left(p_{\varepsilon}, p_{f}, p_{\xi}\right)=p_{\varepsilon} \Delta \varepsilon^{2}+\frac{p_{f}}{3 N} \sum_{i}\left|\Delta \textit{F}_{i}\right|^{2}+\frac{p_{\xi}}{9}\|\Delta \xi\|^{2},
\end{equation}
where $\Delta$ denotes the difference between the prediction of the DP model and DFT results, $N$ is the number of atoms, $\varepsilon$, $\textit{F}_{i}$ and $\xi$ represent the energy per atom, forces of atom \textit{i}, and virial tensor, respectively. $p_{\varepsilon}$, $p_{f}$, and $p_{\xi}$ are tunable prefactors, which are controlled by the learning rate. In this work, the DP model of CIPS was obtained by using the DeepMD-kit software package~\cite{wang2018deepmd,zeng2023deepmd,zeng2025deepmd}, for which the total number of training steps is 10$^7$. The cutoff radius \textit{R$_c$} was set to 11$\textup{~\AA}$, and the smooth cutoff radius was set to 0.5$\textup{~\AA}$ to eliminate the discontinuity introduced by \textit{R$_c$}. The sizes of the embedding neural network and fitting neural network were (25, 50, 100) and (240, 240, 240). During training, the tunable factors $p_{\varepsilon}$, $p_{\xi}$ increased from 0.02 to 1, and $p_{f}$ decreased from 1000 to 1. In addition, the learning rate was reduced exponentially from 0.001 to 1$\times10^{-8}$, and the decay step size was 25000.

The concurrent learning platform DP-GEN, which is based on an active learning strategy, is used to generate datasets~\cite{zhang2019active,zhang2020dp}. Each iteration of this strategy is divided into three steps: 1. \textsl{Training}, 2. \textsl{Exploration}, 3. \textsl{Labeling}. 1. \textsl{Training}: We started from the different polarization configurations of bulk, monolayer, and bilayer CIPS, introduced perturbations to the atomic positions, and manipulated the lattice constants of these configurations through stretching and compressing. Subsequently, we performed ab initio MD (AIMD) simulation at 50 K for these configurations to generate the initial dataset. Using the initial dataset, four DP models were trained. 2. \textsl{Exploration}: We use one of the four DP models obtained in the first step to perform MD simulation using the large-scale atomic/molecular massively parallel simulator (LAMMPS)~\cite{plimpton1995fast,thompson2022lammps}. These MD simulations adopted the NPT ensemble, and temperature and pressure were controlled by a Nose-Hoover thermostat and Parrinello-Rahman barostat, respectively~\cite{parrinello1981polymorphic,nose1984unified}. To explore the configuration space at different temperatures and pressures, eight temperatures (from 100 to 700 K) and three pressures (from 0 to 10 kbar) were set. In addition, for all configurations sampled in the MD trajectory, the maximal standard deviation $\sigma_{f}^{max}$ of the atomic force predicted by the four models was calculated. Through this model deviation, all configurations will be divided into three categories. The configurations of $\sigma_{f}^{max}<\sigma_{low}$, $\sigma_{f}^{max}>\sigma_{high}$, $\sigma_{low}<\sigma_{f}^{max}<\sigma_{high}$ are labeled as accurate, failed and candidate, respectively. In this work, $\sigma_{low}$ and $\sigma_{high}$ were set to 0.05 and 0.20 eV/$\textup{\AA}$. 3. \textsl{Labeling}: These configurations that were labeled candidates in the second step were performed by DFT calculations, and the calculation results were added to the initial dataset, which was used as the training dataset for the next iteration. After 16 rounds of iteration, we finally obtained a dataset with about 18,000 configurations, which contains different polarization configurations of bulk, monolayer, bilayer, trilayer, and quadrilayer CIPS.

All the DFT calculations were performed using the Vienna Ab initio Simulation Package (VASP)~\cite{kresse1996efficient}. The projector augmented wave method was used to construct the pseudopotential~\cite{blochl1994projector,kresse1999ultrasoft}, and the plane-wave energy cutoff value was set to 500 eV. The exchange-correlation functional is parametrized using the Perdew–Burke—Ernzerhof (PBE) formalism in the generalized gradient approximation~\cite{PhysRevLett.77.3865}. The vdW dispersion forces were calculated by the DFT-D3 method~\cite{grimme2010consistent}. The two-dimensional Brillouin zone (BZ) was sampled using a Monkhorst-Pack $k$-mesh with a spacing of 0.16 $\textup{\AA}^{-1}$. A 20$\textup{~\AA}$ vacuum region was used for CIPS thin films to avoid the artificial interaction between neighboring periodic images. The simulation of the phase transition of the bulk CIPS and 3L-CIPS was performed using LAMMPS. We used a 12 $\times$ 6 $\times$ 6 supercell containing 17280 atoms and adopted the NPT ensemble for the bulk simulation. Temperature and pressure were controlled by a Nose-Hoover thermostat and a Parrinello-Rahman barostat, respectively. At each temperature, the simulation time was 15 ns to ensure that the system reached equilibrium, and the step size was 1 fs. The polarization of CIPS was calculated by the atomic coordinates displacement multiplied by the Born effective charge\cite{maisonneuve1997ferrielectric,he2023unconventional}. The Born effective charge of Cu, In, and P$_2$S$_6$ is +0.6, +1.8, and -2.4, respectively, which was used in the previous studies\cite{maisonneuve1997ferrielectric,he2023unconventional}. For the 3L-CIPS, We also adopted the NPT ensemble and selected the FE state as the initial structure. The total simulation time was 1000 ns to ensure capturing the FE-to-FiE transition. The lattice dynamic properties were calculated using the PHONOPY package~\cite{togo2023implementation,togo2023first}.

\section{Results and discussions}
\subsection{DP model}
\label{DP}

\begin{figure}[htbp]
	\centering
	\includegraphics[width=1\linewidth]{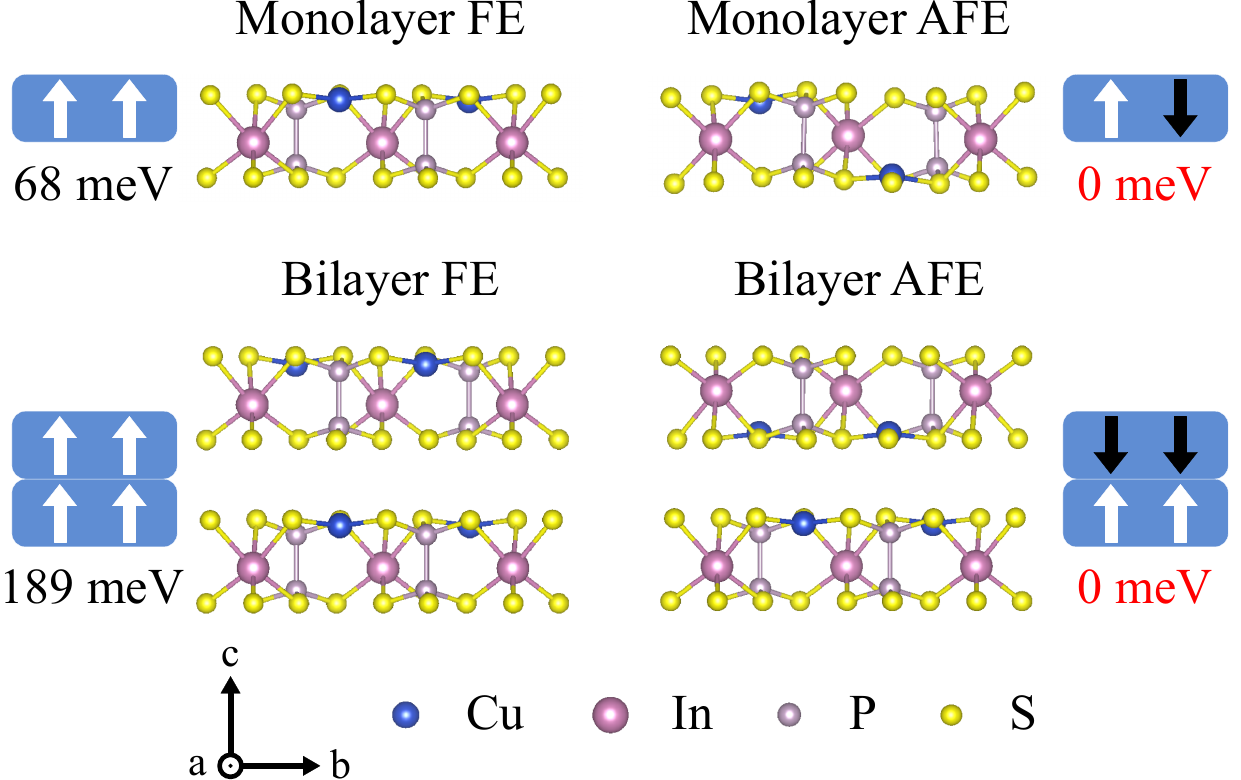}
	\caption{Geometric structures of CIPS monolayer and bilayer with different polarization configurations. The white and black arrows denote the upward and downward polarizations, respectively.}
	\label{fig1}
\end{figure}

We first calculate the energies of different polarization states for monolayer and bilayer CIPS using DFT calculations. We consider two polarization states for the monolayer, i.e., the intralayer FE and AFE orderings (see Fig.~\ref{fig1}). The AFE state is about 68 meV lower than the FE one. For the bilayer, there are eight configurations, which involve both intralayer and interlayer FE and AFE couplings (see Fig.~\ref{fig1} and Appendix~\ref{appa}). Among them, the state containing intralayer FE and head-to-head interlayer AFE orderings has the lowest energy. This state is referred to as the AFE state. Whereas the one with both intralayer and interlayer FE couplings (referred to as the FE state) is about 189 meV higher than the AFE state. Our DFT results are consistent with previous studies~\cite{yu2021interface,reimers2018van,he2023unconventional,sun2019strain,tawfik2018van,yu2023polarization,li2024realization}.

\begin{figure*}[htbp]
	\centering
	\includegraphics[width=1\linewidth]{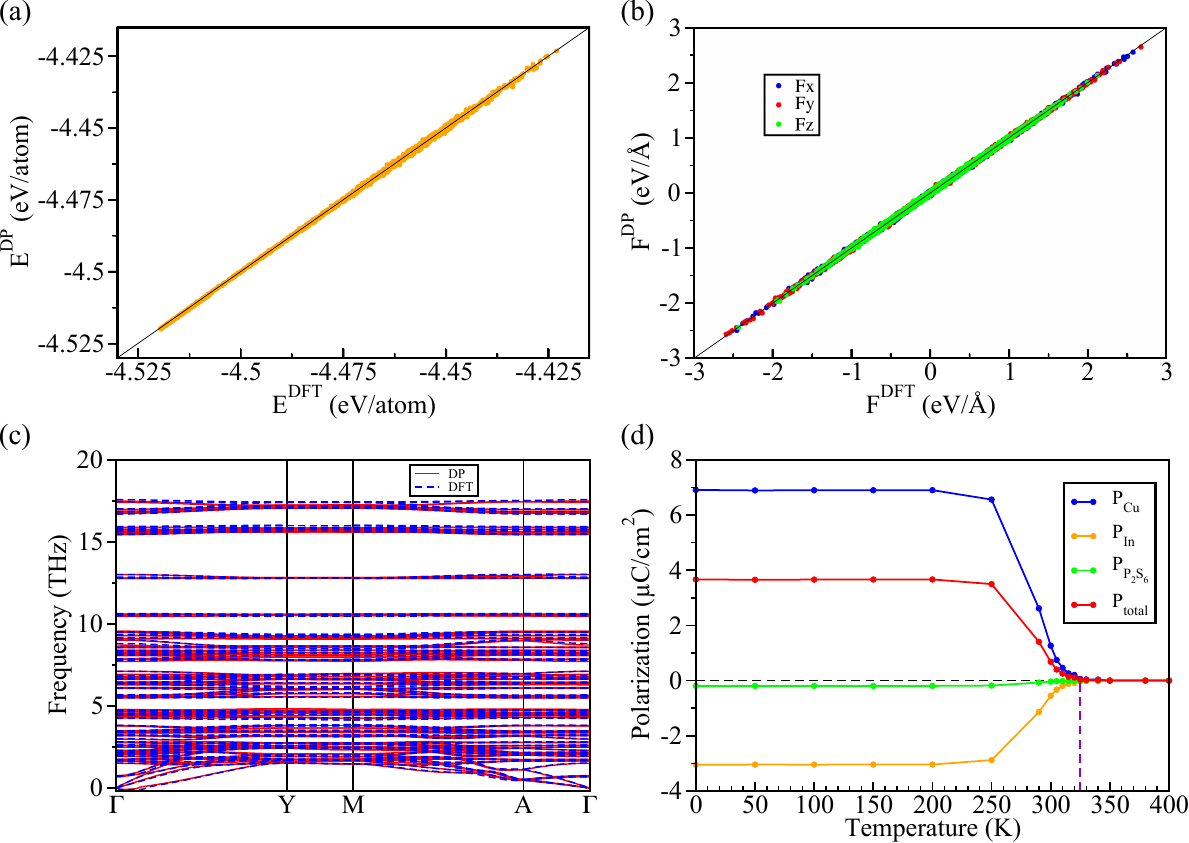}
	\caption{Benchmark calculations of the DP model. Comparison of energies (a) and forces (b) predicted by the DP model and DFT results for all the configurations in the training datasets. (c) The phonon spectrum of the FE state for bulk CIPS from DP (blue dashed line) and DFT (red solid line) calculations. (d) The net polarization of the bulk CIPS as a function of temperature calculated from MD simulations.}
	\label{fig2}
\end{figure*}

To explore the stability of polarization states in multilayers CIPS, we developed a DP model and rigorously validated its accuracy. Figs.~\ref{fig2}(a) and \ref{fig2}(b) show a comparison of the energy and forces predicted by the DP model with DFT results in the training datasets, respectively. The root mean square errors (RMSE) are only about 4.57 $\times~10^{-4}$ eV/atom for energy and 0.023 eV/$\textup{\AA}$ for forces. These results suggest the high accuracy of our DP model. Indeed, the energies calculated by this model for the polarization states shown in Fig.~\ref{fig1} agree well with the DFT results. Moreover, we calculate the phonon dispersions for the FE bulk phase of CIPS using the model, for which the results are shown in Fig.~\ref{fig2}(c). Again, one can see that the results agree well with those derived from the DFT calculations. We also investigated the temperature-driven phase transition of bulk CIPS using DPMD simulations [see Fig.~\ref{fig2}(d)]. The DP model predicts a critical transition temperature T$_{c}$ of 325 K, which agrees with experimental measurement \cite{maisonneuve1997ferrielectric} (315 K) and a previous DP study~\cite{he2023unconventional} (335 K). The above benchmark tests suggest that the obtained DP model is effective in investigating the stability of polarization states in CIPS thin films.

\subsection{Multilayers}
\label{multilayers}
\begin{figure}[htbp]
	\centering
	\includegraphics[width=1\linewidth]{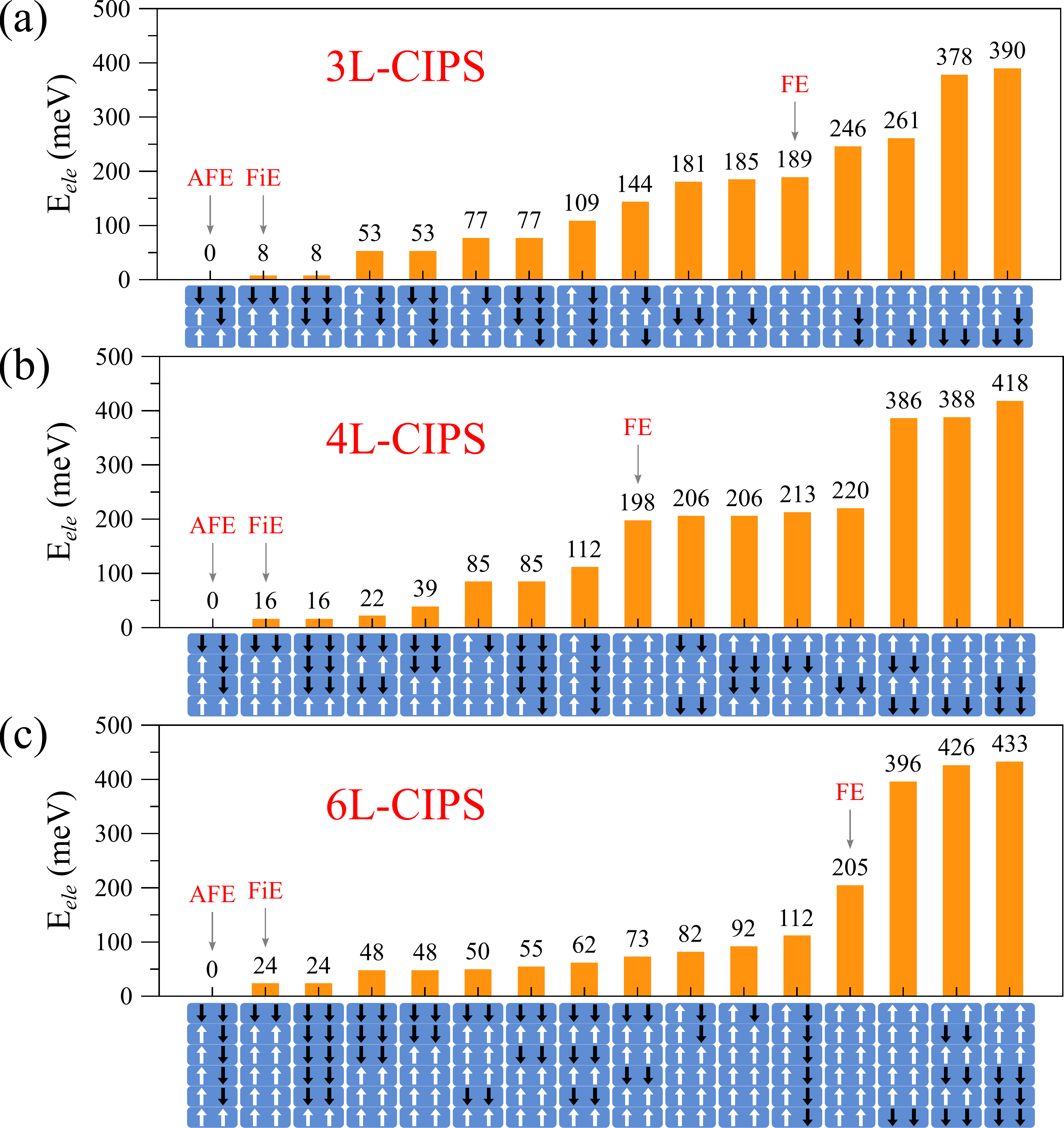}
	\caption{Energetics of various polarization configurations obtained from DFT calculations for 3L$-$(a), 4L$-$(b), and 6L$-$CIPS (c), respectively. The total energy of the AFE state is used as a reference. The polarization configuration is given below each color bar. For 6L$-$CIPS, those with high energies are not shown.}
	\label{fig3}
\end{figure}

Now we apply the DP model to the multilayers of CIPS. For convenience, the trilayer and quadrilayer are denoted by 3L and 4L, respectively. Likewise, for the thin films containing 6, 20, and 40 CIPS layers are represented by 6L, 20L, and 40L, respectively. For 6L$-$CIPS ($\sim$ 3.9 nm), we calculated the energy for all possible polarization configurations (the total number is 4096) using the DP model. Fig.~\ref{fig3} shows the results for a few typical polarization states, identifying the AFE state as the ground state with FE-aligned surface layers (top: downward polarization, bottom: upward polarization) and intralayer AFE coupling for the inner layers. This state has no net polarization. The energy difference between the FE state and the AFE state increases slightly as the thickness of the film grows. 


Note that a ferrielectric (FiE) state has a tiny energy difference from the AFE state. In this state, the surface layers have the same FE ordering as in the AFE state. However, the inner layers exhibit uniform FE ordering. As a result, it has a net polarization comparable to that of the FE state as the thickness of CIPS films grows. The energy differences between the AFE ground state and the FiE state are approximately 8 meV, 16 meV, and 24 meV for 3L$-$, 4L$-$, and 6L$-$CIPS, respectively. Such tiny energy differences suggest that they may be accessed at finite temperatures. In addition to the FiE and AFE states, there are many low-lying polarization states, which have the intralayer FE ordering and partial or full interlayer AFE ordering. 

\begin{figure}[htbp]
	\centering
	\includegraphics[width=1\linewidth]{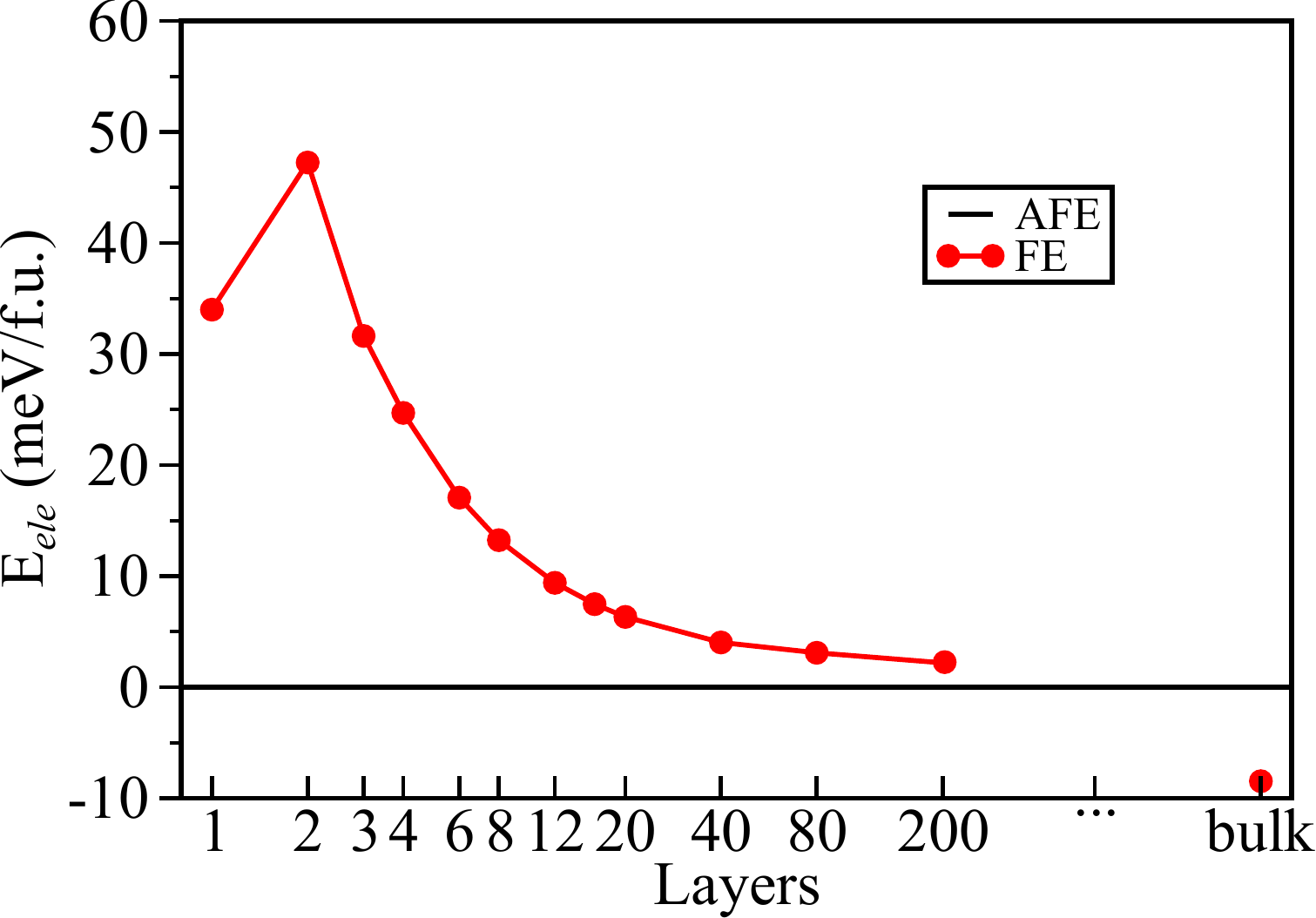}
	\caption{Comparison of the electronic energies of the FE and AFE states for the CIPS thin films as a function of the thickness. The energy of the AFE state is taken as the reference. The configurations of the AFE states for the films in different thicknesses are exactly the same as those shown in Figs.~\ref{fig1} and ~\ref{fig3}.}
	\label{fig4}
\end{figure}

We also performed calculations for thicker films. Fig.~\ref{fig4} compares the energies of the FE state and the AFE state as a function of the thickness. One can see that the AFE state remains more stable than the FE state, even when the thickness of the film is up to 200L. This trend is in contrast to the fact that experimentally the CIPS thin films exhibit net polarizations~\cite{maisonneuve1997ferrielectric,liu2016room}. 

\subsection{Phonon free energy of CIPS}
\label{phonon}
The observed theory-experiment discrepancy motivates rigorous examinations of energy component contributions. While conventional DFT calculations consider only the electronic energy, the thermodynamic ground state at finite temperatures requires inclusion of phonon free energy:
\begin{equation}
E_{total}(T) = E_{ele} + E_{ph}(T)
\end{equation}
Thus, to obtain results that are consistent with experimental measurement at finite temperatures, the contribution of phonon free energy to the total energy may be included in the calculations. Previous studies have shown that the phonon free energy is crucial for the phase stability of crystals~\cite{gu2021significant,ehsan2021first,wolverton2001entropically,colinet2005ab}.

\begin{figure*}[htbp]
	\centering
	\includegraphics[width=1\linewidth]{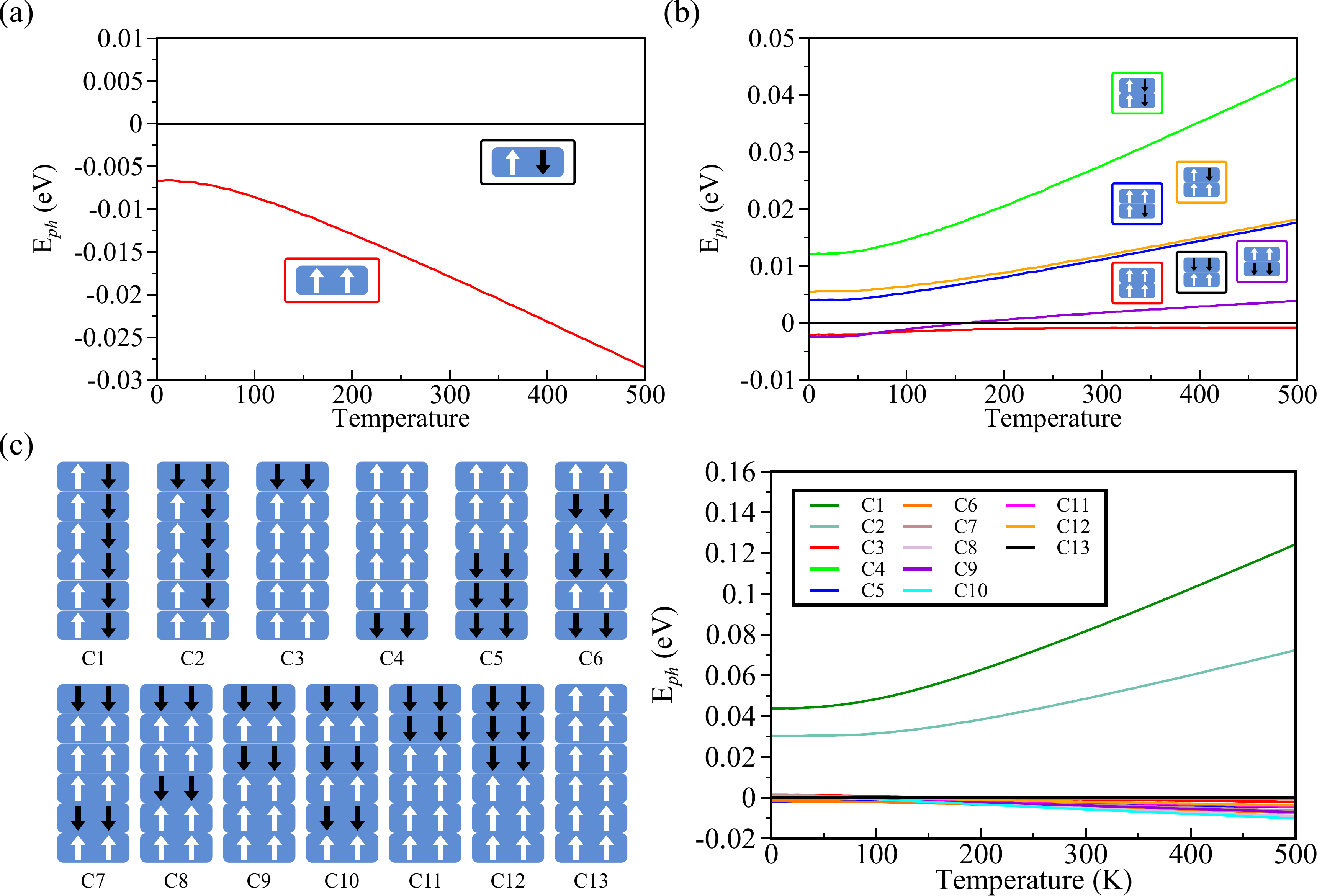}
	\caption{Phonon free energy of CIPS thin films. The phonon free energy of the monolayer (a), bilayer (b), and 6L$-$CIPS (c). For monolayer and bilayer CIPS, the energy of the AFE state is used as the reference. For 6L-CIPS, we use the energy of the FE state as the reference.}
	\label{fig5}
\end{figure*}

We systematically investigate the phonon free energy characteristics for CIPS thin films. Figs.~\ref{fig5}(a) and (b) show the temperature-dependent phonon free energy for various polarization states in monolayer and bilayer CIPS, respectively. Our results show that the intralayer FE state in monolayer CIPS has a lower phonon free energy than the intralayer AFE state. This trend suggests that the stability of the FE state tends to be enhanced at finite temperature for the CIPS monolayer. For the bilayer CIPS, we calculate the phonon free energies for six distinct polarization configurations, which can be categorized into three distinct groups according to their free energies. Group I, exhibiting the intralayer AFE ordering for both layers, has the highest phonon free energy. Group II has three states, which involve intralayer FE configurations only. The energy differences between these states are negligible. However, they have much lower energies than group I. The states of group III contain both intralayer FE and AFE orderings. The phonon free energies for these states are between those for group I and group II. These results suggest that the intralayer polarization configuration plays an important role in the stability of the CIPS thin films, where the intralayer FE configuration becomes favorable at finite temperatures. 

In addition to the monolayer and bilayer, we have also extended the investigations to thicker films. In Fig.~\ref{fig5}(c), we show the results for 6L CIPS. We considered a series of states for this system with different combinations of intralayer FE and AFE orderings. Some of them are shown in the left panel of Fig.~\ref{fig5}(c). The state C1 has the intralayer AFE ordering for all the CIPS layers. C2 has the FE ordering for the surface layer and the intralayer AFE ordering for the inner layers, which has the lowest energy in the DFT calculations (the AFE state shown in Fig.~\ref{fig3}). The others C3-C13 have intralayer FE ordering in every individual layer, but with different interlayer couplings. Fig.~\ref{fig5}(c) shows that C1 has the highest phonon free energy. C2, which has a mixture of FE and AFE ordering, is energetically lower than C1. Although C3-C13 have different net polarizations, their phonon free energies are nearly identical and significantly lower than those of C1 and C2.

Fig.~\ref {fig5} shows that the energy differences between the polarization states are almost identical for each pair of the intralayer FE and AFE orderings at zero temperature, which improve as the temperature increases. For the monolayer, it is 6.7 meV (see fig.~\ref{fig5}a). For the bilayer, the energy differences between group I with full intralayer AFE and group II with full intralayer FE are about 14.2 meV. Therefore, we obtain an energy difference of about 7 meV per pair of intralayer FE and AFE ordering. Likewise, one can also see the trend in other CIPS thin films. Based on this observation, one can expect that the states (e.g., the FiE state) that have an electronic energy close to the AFE ground state may become energetically favorable when the phonon free energy is taken into account.

\begin{figure*}[htbp]
	\centering
	\includegraphics[width=0.9\linewidth]{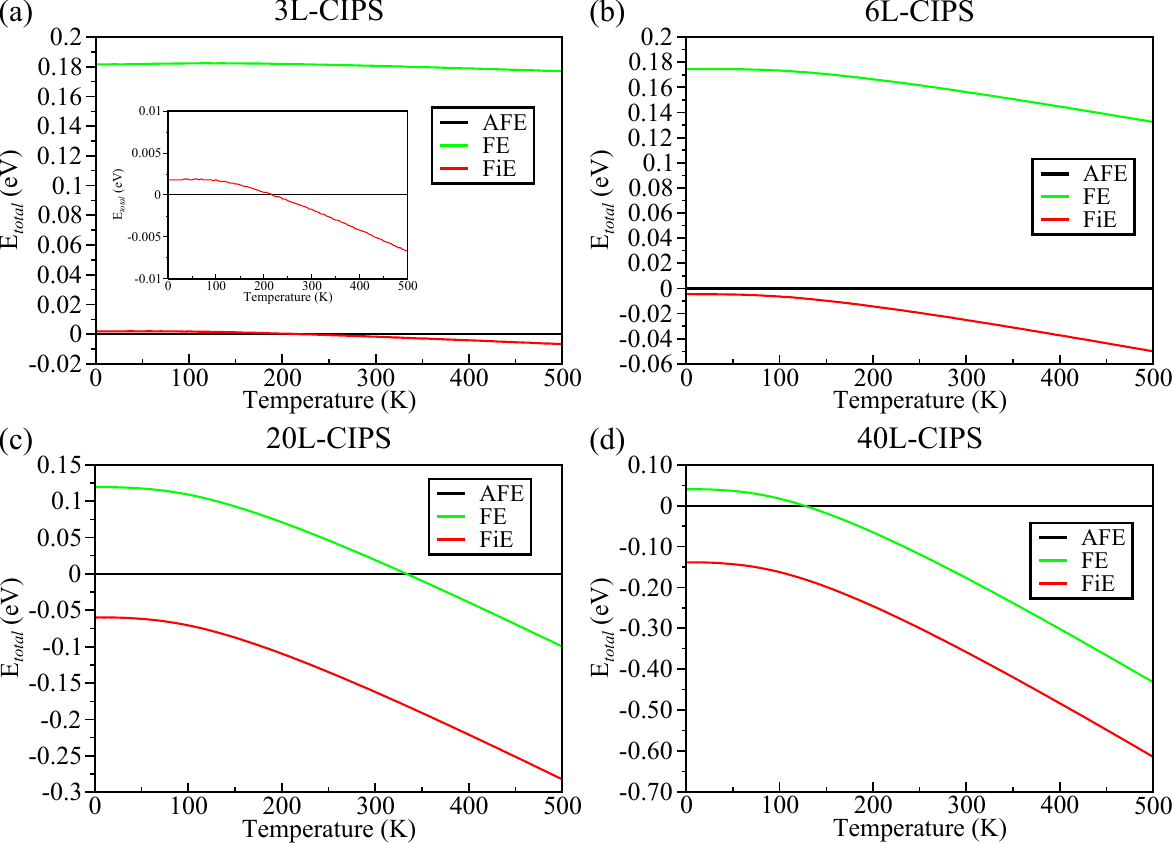}
	\caption{Stability of polarization states in CIPS multilayers. (a) - (b) show the differences in the total energies between the FE, FiE, and AFE states for 3L$-$(a), 6L$-$(b), 20L$-$(c), and 40L$-$(d). The total energy of the AFE state is taken as the reference.}
	\label{fig6}
\end{figure*}

Fig.~\ref{fig6} shows the total energies of the FE, AFE, and FiE states of CIPS thin films with different numbers of layers. One can see that the FE state remains energetically higher than the AFE state for 3L-CIPS. However, the FiE state becomes more stable than the AFE state when the temperature is higher than 220 K. We used the DP model to simulate the dynamic evolution of 3L-CIPS at 250 K. We selected the FE state as the initial structure. The FE state transformed into the FiE state after a long simulation (1000 ns), which further confirmed our conclusions. For 6L-CIPS and thicker CIPS films, the FiE state is more stable than the AFE state throughout the whole temperature range. Note that for 20L-CIPS and 40-CIPS, the FE state also has a lower energy than the AFE state above a certain temperature. However, it remains to be energetically higher than the FiE state. This result further supports a recent experimental observation that CIPS thin films show intralayer FE ordering and interlayer AFE coupling between the surface layer and the underneath layer~\cite{neumayer2022nanoscale}. Additionally, we have employed the DP model from a previous study to calculate the total energies of the three polarization states in 6L-CIPS~\cite{he2023unconventional}. Although there is a light difference in the predicted transition temperatures between our model and the one by Ref.~\cite{he2023unconventional}, both models show the same trend that the FiE state has a higher stability than the AFE state at certain finite temperatures.

\begin{figure}[htbp]
	\centering
	\includegraphics[width=1\linewidth]{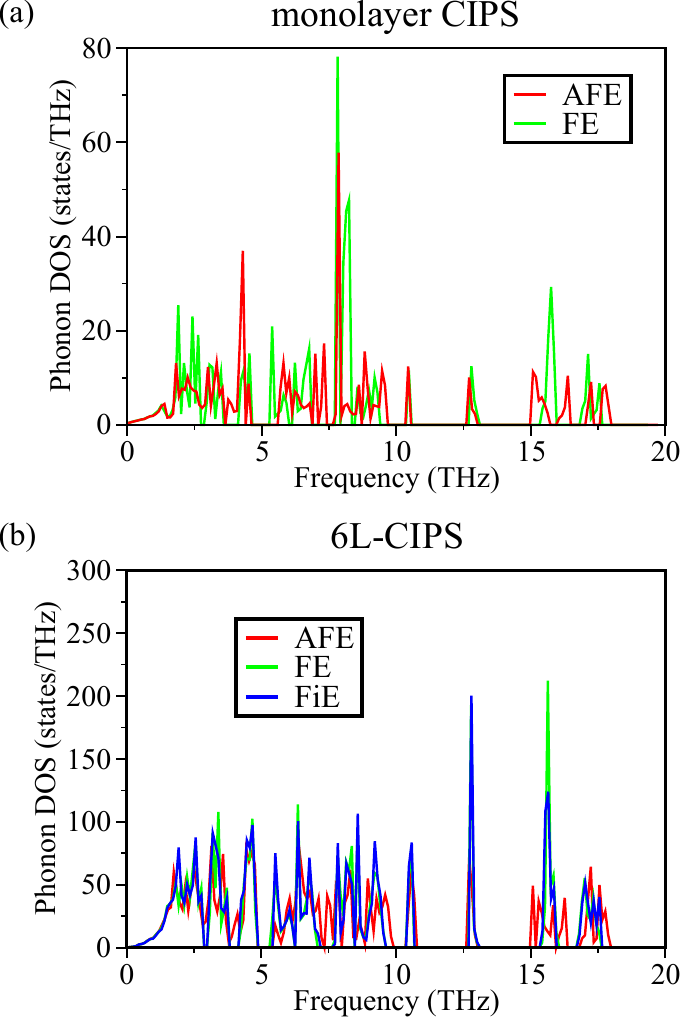}
	\caption{Phonon DOS of different polarization configurations for monolayer (a) and 6L$-$(b) CIPS.}
	\label{fig7}
\end{figure}

The difference in the phonon free energy between the FE, FiE, and AFE states can be understood by an analysis of lattice vibrations in CPIS thin films. Fig.~\ref{fig7} shows the phonon density of states (DOS) of different polarization states for monolayer CIPS and 6L$-$CIPS. For the monolayer CIPS, the FE ordering has a higher phonon DOS in the low energy region than the AFE state, which gives rise to a lower phonon free energy for the FE state. For the 6L-CIPS, the phonon DOS for the FE state and the FiE state are pretty much similar, which explains why their phonon free energy are comparable. The FE state only has a higher phonon DOS than the FiE state at about 16 THz. Therefore, the phonon free energy for the FE state is slightly higher than that of the FiE state. The phonon DOS of the AFE state shows upshifts in the frequency located in the region of 5$-$20 THz relative to the FE and FiE states, leading to a higher phonon free energy for the AFE state. These results establish a direct polarization-vibration-energy relationship for the various polarization states in CIPS thin films.

\section{Conclusions}
\label{conclusions}
In summary, through systematic first-principles calculations and the DP method, this study elucidates the critical role of phonon thermodynamics in governing polarization phase stability of CIPS thin films. We find that CIPS thin films are in the AFE state based on the DFT electronic energy, which is consistent with previous theoretical reports. Then, a high-precision DP model was developed and applied to the lattice dynamics of CIPS in various polarization states. Our results reveal a fundamental mechanism: the phonon free energy of CIPS thin films depends on the intralayer polarization ordering, and the intralayer FE ordering shows a lower phonon free energy than that of intralayer AFE ordering. The interlayer polarization coupling has a minor effect on the phonon free energy. In addition, we have found a metastable FiE state with intralayer FE ordering that exhibits competitive electronic energy with the AFE state. By including the contribution of the phonon free energy, the FiE state shows a higher stability than the AFE state. Our study thus reconciles the long-standing discrepancy between theoretical predictions and experimental observations on the ground state of CIPS thin films. 

\begin{acknowledgments}
This work was supported by the National Natural Science Foundation of China (Grants No. 12574262, No. 12174098 and No. 52372260), the Major Fundamental Research Program of Hunan Province (2025ZYJ004), the Science Fund for Distinguished Young Scholars of Hunan Province of China (Grant No. 2024JJ2048), the Youth Science and Technology Talent Project of Hunan Province (Grant No. 2022RC1197). Calculations were carried out in part using computing resources at the Hunan Normal University High Performance Computing Platform. 
\end {acknowledgments}

\section*{DATA AVAILABILITY}
The data that support the findings of this article are not publicly available upon publication because it is not technically feasible, and the cost of preparing, depositing, and hosting the data would be prohibitive within the terms of this research project. The data will be available from the authors upon reasonable request.

\appendix
\section{Polarization configurations for bilayer CIPS}
\label{appa}

\begin{figure}[htbp]
	\centering
	\includegraphics[width=0.8\linewidth]{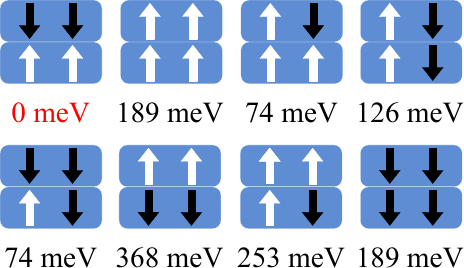}
	\caption{Polarization configurations for bilayer CIPS}
	\label{fig8}
\end{figure}

We consider 8 different polarization configurations for bilayer CIPS. Fig.~\ref{fig8} shows their corresponding energy, and the ground state of the bilayer CIPS is the AFE state.

\section{The total energies of 6L-CIPS calculated by a different DP model}
\label{appb}

\begin{figure}[htbp]
	\centering
	\includegraphics[width=1\linewidth]{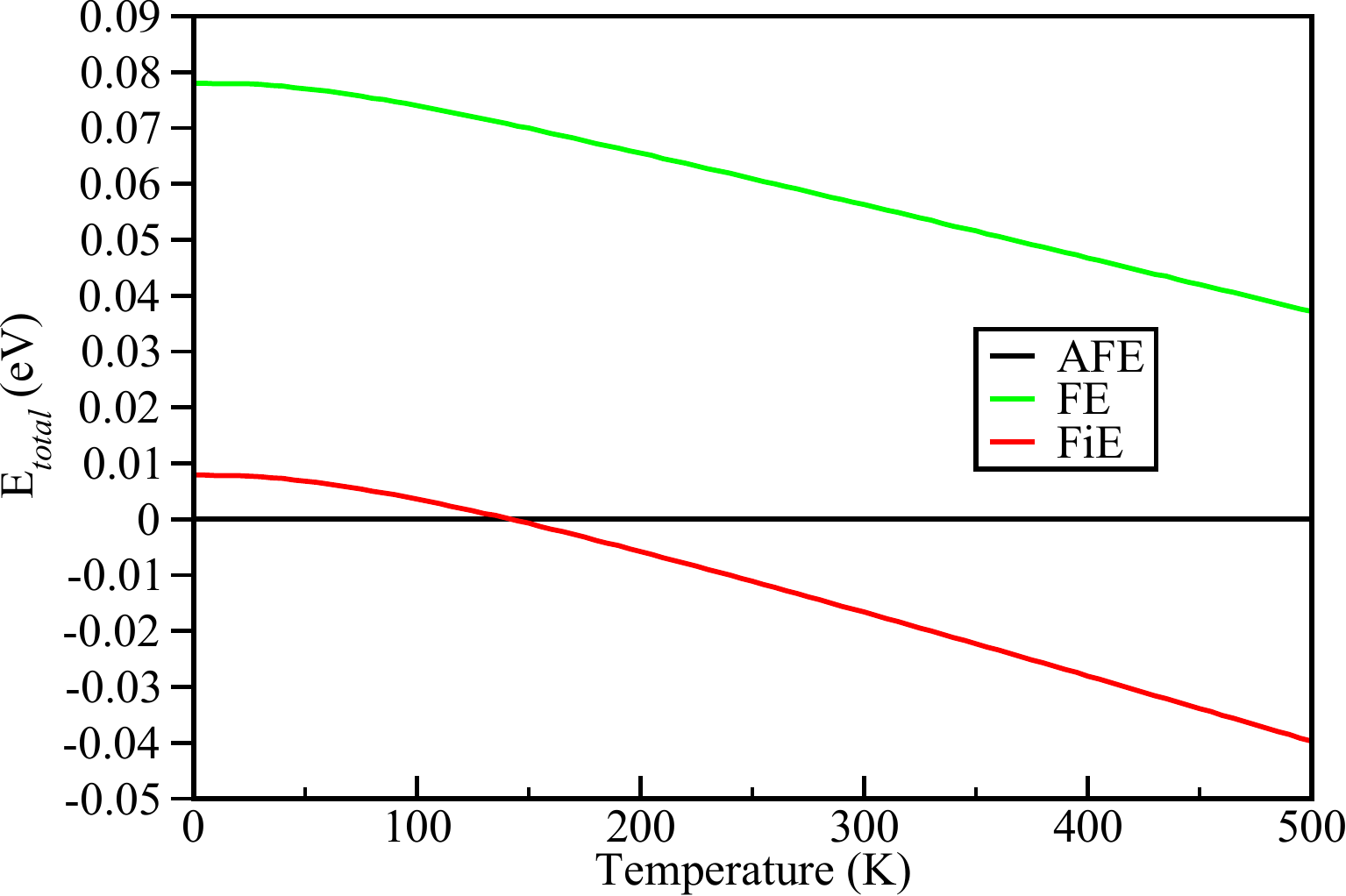}
	\caption{Total energies of 6L-CIPS at finite temperatures calculated by the DP model from Ref.~\cite{he2023unconventional}.}
	\label{fig9}
\end{figure}

We have also calculated the total energy by including the phonon free energy using the DP model from a previous study~\cite{he2023unconventional}. The results summarized in Fig.~\ref{fig9} show the same trend as our DP model.

\section{The energies of different polarization states for 6L-CIPS.}
\label{appc}

In Table~\ref{table1}, we comprehensively compared the electronic energy, phonon free energy, and the total energy of the 13 polarization states shown in Figure 5(c) at 300 K. The results clearly demonstrate that the phonon free energy plays a crucial role in the stability of different systems.
\begin{table}[b]
\caption{\label{table1}%
A comparison of the electronic energy, phonon free energy, and total energy of the 13 polarization states shown in Figure 5(c) at 300 K. For each type of energy, the lowest one is used as the reference.
}
\begin{ruledtabular}
\begin{tabular}{cccc}
\textrm{States}&
\textrm{E$_{ele}$ (meV)}&
\textrm{E$_{ph}$ (meV)}&
\textrm{E$_{total}$ (meV)}\\
\colrule
C1 & 112 & 88 & 171\\
C2 & 0 & 55 & 26\\
C3 & 24 & 5 & 0\\
C4 & 396 & 5 & 372\\
C5 & 433 & 3 & 407\\
C6 & 426 & 1 & 398\\
C7 & 50 & 3 & 24\\
C8 & 73 & 1 & 45\\
C9 & 55 & 2 & 28\\
C10 & 62 & 0 & 33\\
C11 & 48 & 4 & 23\\
C12 & 48 & 4 & 23\\
C13 & 205 & 6 & 182\\
\end{tabular}
\end{ruledtabular}
\end{table}
\color{black}

\bibliography{references}
\end{document}